\documentclass[5p]{elsarticle}
\usepackage[UKenglish]{babel}
\usepackage{graphicx}
\usepackage{dcolumn}
\usepackage{bm}
\usepackage{epsfig,amsmath,mathrsfs}
\usepackage{amssymb}
\usepackage{xcolor,hyperref}
\hypersetup{final}

\def \nn{\nonumber}
\def \tl{\tilde}

\journal{Physics Letters B}

\begin{document}

\begin{frontmatter}

\title{Hawking radiation from acoustic black holes in relativistic 
heavy ion collisions}

\author[prl,pol]{Arpan Das}
\ead{arpan.das@ifj.edu.pl}

\author[imsc]{Shreyansh S. Dave}
\ead{shreyanshsd@imsc.res.in}

\author[prl]{Oindrila Ganguly\corref{cor1}\fnref{fn1}}
\ead{oindrilag@iisc.ac.in}

\author[iop]{and Ajit M. Srivastava}
\ead{ajit@iopb.res.in}

\cortext[cor1]{Corresponding author}
\fntext[fn1]{Present affiliation: Indian Institute of Science, Bengaluru 560012, India}

\address[prl]{Physical Research Laboratory, Ahmedabad 380009, India}
\address[pol]{Institute of  Nuclear  Physics,  Polish  Academy of  Sciences,  PL-31-342  Krak\'ow,  Poland}
\address[imsc]{The Institute of Mathematical Sciences, Chennai 600113, India}
\address[iop]{Institute of Physics, Bhubaneswar 751005, India}

\begin{abstract}
We propose a new analogue model of gravity - the evolving quark gluon 
plasma (QGP) produced in relativistic heavy ion collisions. This quark gluon 
plasma is the ``most inviscid" fluid known. Such low kinematic viscosity is
believed to reflect strongly correlated nature for QGP in these
experiments. Hence, it may provide a good example of a quantum fluid 
naturally suited to studies of acoustic Hawking radiation. Due to rapid 
longitudinal expansion, presence of a sonic horizon is also naturally
guaranteed here, though, in general, this horizon is not static. Using \textit 
{Ultra relativistic quantum molecular dynamics} (UrQMD) simulations, we 
show that, under certain conditions, the longitudinal velocity of the 
plasma, near the sonic horizon, can become  time independent for a short 
span during the evolution of the system. During this period, 
we can have a conformally static acoustic metric with a (conformal) Killing 
horizon coinciding with the apparent horizon. An asymptotic observer will 
then see a thermal flux of phonons, constituting the Hawking radiation, 
coming from the horizon. For the relatively low energy collision 
considered here, where the resulting QCD system is governed by non-relativistic 
hydrodynamics, we estimate the Hawking temperature to be about 4-5 MeV
(with the temperature of the QCD fluid being about 135 MeV).
We discuss the experimental signatures of this Hawking radiation in terms 
of a \textit{thermal} component in the rapidity dependence of the transverse 
momentum distribution of detected particles. We also discuss extension to
ultra-relativistic case which should lead to a higher Hawking temperature, 
along with the effects of dynamical horizon leading to blue/red shift of 
the temperature.
\end{abstract}

\end{frontmatter}

\section{Introduction} \label{sec:intro}

One of the startling and not classically intuitive characteristics of a 
black hole is its evaporation. Hawking had shown that black holes 
\textit{spontaneously} emit thermal radiation, named after him as Hawking 
radiation, at a temperature $T_H = 1/ 8 \pi M$ (in natural units), $M$ being 
the mass of the black hole \cite{hawking1975}. For a stellar mass black hole 
of mass $M \approx M_\odot \approx 10^{30}kg$, the Hawking temperature turns 
out to be $T_H \approx 10^{- 7} K$, which is much less than the temperature 
of cosmic microwave background radiation. So, at present stage, it is 
virtually impossible to detect Hawking radiation through astronomical 
observations. 

Hawking radiation is an artefact of the way quantum fields behave in 
curved spacetime. Interestingly, it so happens that in an inviscid fluid 
with barotropic equation of state and irrotational bulk velocity, acoustic 
perturbations in the velocity 
potential obey an equation which is identical to the Klein Gordon equation 
satisfied by a massless scalar field in curved spacetime \cite{unruh1981}. 
Thus, the acoustic perturbations perceive an \textit{effective} acoustic 
spacetime whose geometry is determined  by the bulk velocity profile,  
density and pressure of the fluid.  If the fluid flows in such a way that 
there exists a surface on which at every point the inward normal component of 
the fluid velocity becomes equal to the speed of sound in the fluid and 
becomes supersonic beyond it, then all acoustic perturbations originating 
in the supersonic region are swept inwards by the flowing fluid. This 
surface acts as a horizon for acoustic perturbations, thus forming an 
`acoustic black hole' or a `dumb hole'. Unruh chose a spherically symmetric, 
stationary, convergent background flow to construct an analogue of the 
spacetime outside a Schwarzschild black hole and showed that quantised 
acoustic perturbations could be emitted from the horizon of such a dumb 
hole with a thermal spectrum, just like the emission of radiation from a 
black hole horizon via Hawking radiation \cite{unruh1981,visser1998}. A 
disconcerting aspect of this remarkable find is that it requires the 
quantisation of linearised acoustic perturbations in a classical fluid! In 
reality, to see Hawking radiation in alternate systems, we need a quantum 
analogue model that can be described in terms of a  classical effective 
background spacetime with some standard relativistic quantum fields living 
on it \cite{barcelo2005}. Quantum fluids like superfluid helium, Bose 
Einstein condensates (BEC), photon fluid etc. are promising in this regard 
(for a review, see ref. \cite{barcelo2005, novello2002book}). In fact, very 
recently,  correlation between Hawking particles and their partners beyond 
the acoustic horizon of analogue black holes have been observed in a series 
of experiments using one dimensional flowing atomic condensates 
\cite{steinhauer2015,steinhauer2016,denova2018}. 

A novel system which is yet to be explored from this perspective is the 
longitudinally expanding quark gluon plasma (QGP) created  
in relativistic heavy 
ion collisions (HIC).  Here, two heavy nuclei, moving at speeds close to the 
speed of light, $c$, collide and  move through each other. Owing to the 
extremely high temperature and density produced during collision, a quark 
gluon plasma is formed. This plasma fills up the space between the two 
receding nuclei which continue to move away from each other at speeds 
approaching the speed of light, $c$.  The velocity of 
the plasma fluid monotonically
increases from zero at the centre of collision (in the centre of mass frame) to 
a value very close to $c$ near the receding nuclei. It is then obvious that 
the fluid velocity becomes supersonic at some surface within the plasma 
leading to the formation of an acoustic horizon. A very important observation 
regarding the QGP produced in relativistic heavy-ion collisions at
very high energies (e.g. at RHIC and at LHC) is that the produced
QGP is the ``most inviscid" fluid known with the lowest value of
shear viscosity to entropy density ratio, $\eta/s$, of all known fluids. This 
was a very surprising result, as the expectation was that such a QGP system 
would be close to the ideal gas limit with corresponding large 
viscosity. Such a low value of kinematic viscosity is believed to reflect the 
strongly correlated nature of QGP in these experiments and is referred to
as sQGP (strongly correlated QGP). Hence, this QGP system may provide a good 
example of a quantum fluid, fulfilling all the requirements for 
the presence of acoustic Hawking radiation. Additionally, this 
gravitational analogy can unravel facets of the plasma that have 
remained hidden otherwise. 

We mention that the actual system we study in this paper is a 
non-relativistic plasma of nucleons resulting from an Ultra relativistic quantum 
molecular dynamics (UrQMD) simulation of 
relatively low energy heavy-ion collisions as appropriate for the Beam Energy 
Scan at RHIC, and for FAIR/NICA energies. For such a system, $\eta/s$ is 
expected to be larger, so presence of strong quantum 
correlations is not so obvious. However, as we will show below in Sect. 3, 
even here one can argue for the quantum nature of the fluid from the fact 
that Fermi-Dirac distribution with large chemical potential provides a much 
better fit of the simulation results than  the Maxwell-Boltzmann
distribution. Thus, quantum statistics plays an important role here.
Before we specialise to such a non-relativistic plasma of nucleons  
produced in a low energy collision, it is worthwhile to have a general picture
of the acoustic black hole for the ultra-relativistic case as appropriate for
RHIC and LHC.

For ultra-relativistic collisions, the resulting QGP system is
expected to follow Bjorken's longitudinal boost invariant expansion model
during early stages. Here, flow velocity has only z-component $v^z$ 
(along the beam axis), with the scaling law $v^z = z/t$ as measured in the
centre of mass frame. $t = 0$ corresponds to the instant when the two
(highly Lorentz contracted) nuclei overlap. Subsequently, the nuclei
go through each other, and while receding away at ultra-relativistic speeds, 
populate the intermediate
region with QGP resulting from interactions of partons in the colliding
nuclei. (It has been argued in ref. \cite{kharzeev2005} that acceleration of 
initial partons in the color glass condensate field during this stage 
will lead to Hawking-Unruh thermal radiation which can result in rapid 
thermalisation.) 
The flow velocity of the plasma exceeds the sound velocity at some 
location in the region between each receding nucleus
and the origin of the centre of mass frame,
which becomes the location of the acoustic horizon. 

 However, it is easily seen
that with  Bjorken longitudinal scaling expansion,  $v^z = z/t$,
the acoustic horizon is not static, rather it moves away from the centre
of  collision at the speed of sound.
This is not the usual picture of acoustic black hole where horizon
is supposed to be static, leading to the standard picture of Hawking
radiation. The situation of dynamical horizon for black holes has
deep conceptual issues, and has been extensively discussed in the literature 
(see for example ref. \cite{nielsen2008} and references therein).
Intuitively one may expect that a horizon receding away from the asymptotic
observer should lead to red-shifted Hawking radiation while a horizon
moving towards the observer will lead to blue shifted radiation. For
an acoustic black hole horizon which is moving away with speed of sound
(as in  Bjorken's scaling expansion model), one will then expect
infinite redshift, making the Hawking radiation unobservable.

However, Bjorken's scaling expansion model is not strictly valid for the entire
plasma region, even for ultra-relativistic case. For large rapidity regions,
 especially near the receding nuclei, one expects significant deviations from
 the scaling law due to non-trivial gradients of energy density and pressure. 
 Certainly, for low energy collisions, the scaling law is not
 valid even in the central regions. In such situations, the velocity of the
 acoustic horizon may be much smaller than the speed of sound.
 In fact, as we will see below for simulations with low energy collisions,
 in certain cases the acoustic horizon may even move towards the asymptotic 
 observer located at the centre,
 instead of receding away. Additional richness in the motion of acoustic
 horizon can result from non-trivial dependence of the speed of sound on
 energy density and pressure. To allow for these different possibilities,
 one has to consider the issue of Hawking radiation with dynamical horizon.
 This being a complex issue, we postpone its discussion for future work.
 In the present work we confine our attention to specific situations
 in which, due to a balance between the decreasing velocities at
 a given point due to longitudinal expansion, and acceleration  due to
 non-trivial pressure gradients, one is able to achieve almost static
 acoustic horizon for few fm duration of time. This becomes a conceptually
 clean case where the conventional calculations of Hawking radiation can
 be employed. This situation is achieved in simulations at relatively low
 collision energies, and naturally we find the resulting Hawking temperature
 to also have a small value, about 4 MeV (with the fluid temperature being
 about 135 MeV). Though, this is  a small value,
 and may be difficult from observational point of view, this case
 illustrates the existence of this novel phenomenon of
 Hawking radiation from acoustic black holes in heavy-ion collisions.
 We expect much higher temperatures for the ultra-relativistic case,
 along with effects of red/blue shift of the Hawking radiation due to
 dynamical horizon.

We start with a brief overview of our system in the following section and 
show how an effective dynamical acoustic metric may be obtained under 
certain simplifying assumptions.  In section \ref{sec:urqmd}, we 
discuss conditions for local thermodynamic equilibrium
for plasma with 
UrQMD simulations and then determine the 
space and time dependence of its longitudinal velocity.  We find
a small window of about $2$ fm time during 
which the longitudinal velocity of the plasma becomes almost 
time independent in the region around the 
sonic horizon. Confining ourselves to this tiny window, 
in section \ref{sec:sgrhr}, we write down the corresponding acoustic metric 
which is static but for a spacetime dependent conformal factor. We deduce 
the surface gravity of the (conformal) Killing horizon and the Hawking 
temperature that an asymptotic observer (located at $z = 0$ where the
fluid velocity is zero) would measure. 
We discuss the issue of non-trivial conformal factor, and its time 
dependence, on the Hawking temperature. 
Finally,  in section \ref{sec:obs}, we discuss 
observational prospects of acoustic Hawking radiation in this new analogue 
model of gravity and conclude by pointing out scopes for improvement of 
this first study, in particular, extension to the relativistic hydrodynamics
case and consideration of dynamical horizon. A comment on notation: we use 
Greek alphabets to denote spatial indices and lowercase Latin alphabets to 
denote spacetime indices. 

\section{The system} \label{sec:sys}
In a relativistic heavy ion collision, two heavy nuclei 
(like Pb, Au) approach each other at relativistic energies, along a line, 
collide and pass through each other. 
Owing to the high 
energy of collision, a quark gluon plasma (QGP) is formed (or a hadronic 
plasma, at low energies), after attaining local thermodynamic equilibrium. 
For ultra-relativistic collisions, the  resulting quark gluon plasma (e.g. 
at RHIC and at LHC) is the 
most ideal fluid known to exist in nature, with a value of kinematic viscosity 
($\nu = \frac{\eta}{s}$,  $\eta$ and $s$ being the coefficient of shear 
viscosity and specific entropy of the plasma respectively), 
lower than any other known material in nature \cite{romatschke2007}. 
This is one of the most important, and surprising, results from these 
experiments and is deduced from the measurements of elliptic flow in
non-central collisions \cite{romatschke2007}.

\begin{figure}
\centering
\includegraphics[width=0.85\linewidth]{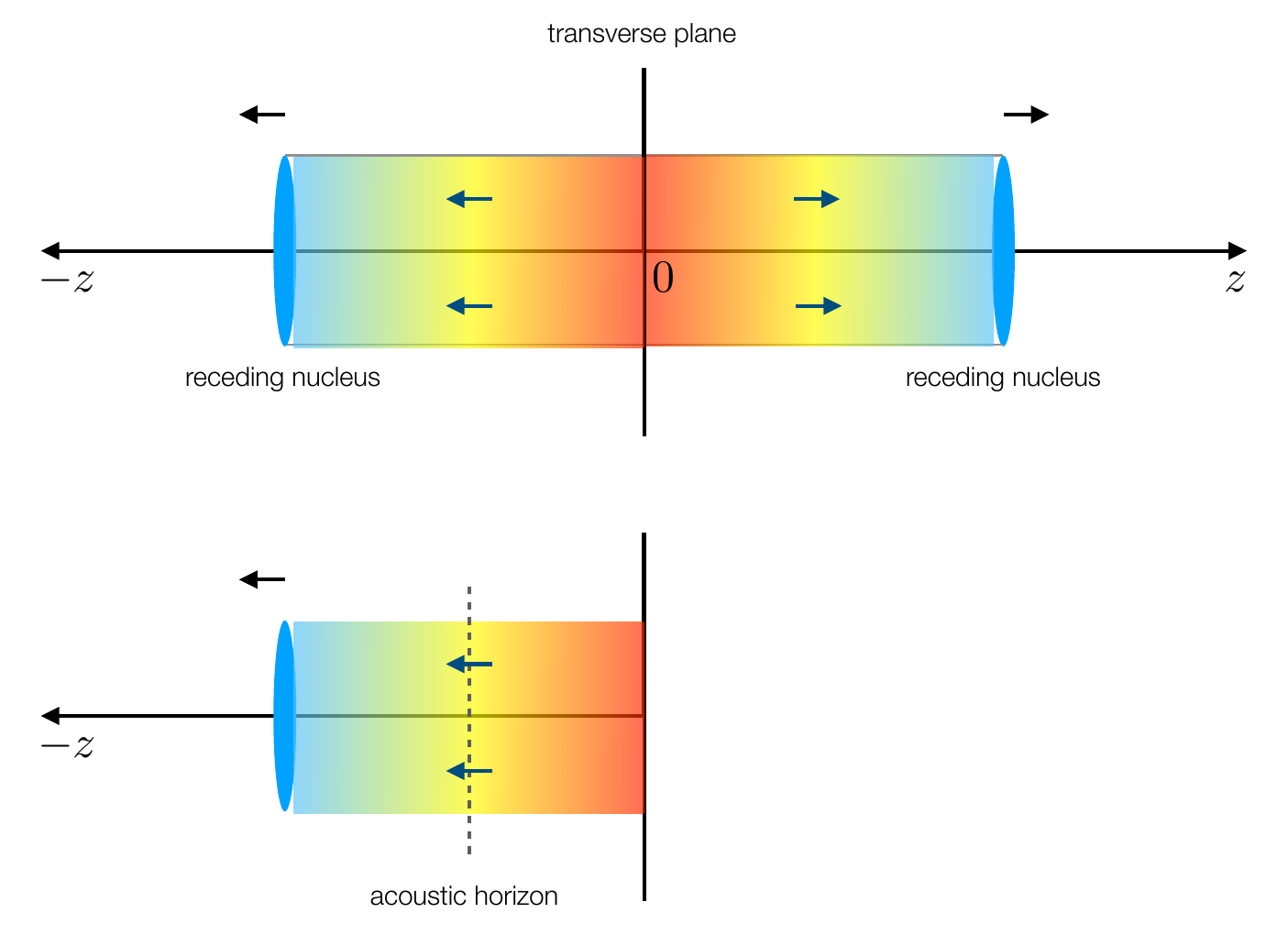}
\caption{\label{fig:coord} Schematic representation of quark gluon plasma 
formed by the collision of two heavy (Lorentz contracted) nuclei. 
For illustration purpose,
longitudinal direction is exaggerated. For the time scales relevant,
longitudinal dimension of the system is much smaller than the transverse 
dimension.}
\end{figure}

The plasma initially acquires only a longitudinal velocity from the 
colliding nuclei and has zero transverse velocity 
(transverse to the beam direction). At this point, to visualise the 
scenario, we introduce a 
laboratory observer whose frame coincides with the centre of mass (CoM) 
frame of the colliding nuclei with the  $z$ axis along
the longitudinal direction. The time of collision is taken to be
$t=0$. 
Figure \ref{fig:coord} schematically represents the situation at a 
particular  instant of time after collision. Initially, the three velocity 
of the plasma has the form 
$v^\alpha (t,x,y,z) = (0,0,\pm v^z (t,z))$ where the upper and lower 
signs denote velocities of the plasma in the right $(z > 0)$ and left 
($z< 0$) half spaces respectively. $\lvert v^z \rvert$ ranges from zero at 
$z=0$ to almost the speed of light close to the receding nuclei. Let 
$\epsilon (t,x,y,z)$ and $p(t,x,y,z)$ be the energy density and pressure 
of the plasma respectively, related by a 
barotropic equation of state (EoS) $p \propto \epsilon$ 
\cite{philipsen2012,borsanyi2010,borsanyi2012,parotto2018,bazavov2017}. 
Now, we make three simplifying assumptions to be employed while 
viewing the system as an analogue model of gravity. Firstly, we assume that 
the velocities $v^x, v^y$ can be ignored in comparison to $v^z$ 
for the time scales considered here. This is expected for times less
than $\sigma_T/c_s$ \cite{ollitrault2007}
 where $\sigma_T$ is the
initial transverse size of the resulting plasma (expected to be of order 3
fm for a central Au-Au collision at RHIC energies), and $c_s$ is the speed
of sound. (This is a rough estimate. For a more careful account of the
transverse expansion time scale, one should use better estimates, such
as using the Blast wave model \cite{teaney2000}.
We are not able to address such details, this being a preliminary investigation
of this kind. Also, in our way of determining equilibrium in a relatively
large cell at each value of $z$, to some extent, any transverse expansion would
already have been accounted for. We can probably safely say that at least
within roughly the region of  transverse dimensions (about 5 fm) of the
cells used for determining equilibrium, one can neglect transverse expansion
for the relevant time scales here.)
Thus the variables $\vec v, \epsilon, p$ will be 
functions only of time $t$ and position along the $z$ axis. 
Thus, the velocity field can now simply be written as $v^\alpha (t,z) = 
(0,0,\pm v^z (t,z))$.  The monotonic increase of $\lvert v^z \rvert$ 
from zero to values close to the speed of light in each half space ensures 
that somewhere in each half of the expanding plasma, $\lvert v^z \rvert$ 
crosses the sonic value and an acoustic horizon is formed. 

The dynamics of 
the quark gluon plasma should be described by relativistic equations for the 
conservation of energy and momentum as the plasma velocity takes relativistic 
values and its equation of state is that for a relativistic fluid. However, 
as a first approximation towards constructing a new analogue model of gravity, 
in this work we use non-relativistic hydrodynamics 
instead. This is not unreasonable, as for the actual QGP system, we will
restrict to  fluid velocities of order, and less than, the sound speed
which will be less than $1/\sqrt{3}$ (in units of speed of light, $c$). 
Thus, relativistic corrections will
not be very large. For the equation of state part, we are only assuming
a barotropic equation of state, so the relativistic aspect of equation
of state is not playing a major role here.  In fact the UrQMD
simulation we have carried out, deals only with hadronic degrees of freedom,
and we include only nucleons (protons and neutrons), so our non-relativistic 
treatment is consistent 
with that.  Under this approximation, the mass density  $\rho(t,z)$  of the 
plasma becomes relevant in place of the energy density $\epsilon (t,z)$ and 
the barotropic condition is $\rho(p)$.  
We consider the stage when thermal equilibrium has been attained
(to a good approximation, as discussed below) so that nucleons follow a 
Fermi Dirac distribution with a large value of the baryon chemical potential, 
(instead of the  Boltzmann velocity distribution which holds for longer
times \cite{bravina1998}). Thus quantum statistics is relevant, again 
justifying the quantum nature of the fluid. 
This observation is very important. Note that we earlier
argued that for ultra-relativistic collisions, the very low values of 
$\eta/s$, close to the AdS/CFT limit, suggest presence of strong quantum 
correlations, justifying the quantum nature of fluid which is of crucial
importance for Hawking radiation from the acoustic black hole. However,
for the non-relativistic nucleonic plasma resulting from a low energy 
heavy-ion collision, $\eta/s$ is larger, so its value cannot be used as
an indicator of the presence of strong quantum correlations. However, as 
we discussed above, the quantum nature of fluid here is manifest from
the fact that quantum statistics with the
Fermi-Dirac distribution provides a much better fit  compared to 
the Maxwell-Boltzmann distribution.

When talking about the acoustic spacetime, we intend to restrict ourselves 
to the region $- \infty < z \leq 0$. 
This ensures that the plasma flows towards decreasing values of the
longitudinal coordinate $z$. The velocity field 
$v^\alpha (t, z) = (0,0,- v^z (t,z))$ is 
naturally irrotational, thus allowing us to express the velocity 
$\vec{v}(t, z)$ in terms of a scalar velocity potential 
$\psi (t, z)$ as $\vec{v} = \vec{\nabla} \psi$. 

With these assumptions in place, the plasma satisfies all the criteria 
necessary for the construction of an analogue model of gravity. Hence, we 
can directly write down the effective acoustic metric emerging in this 
plasma, in coordinates $(t,x,y, z)$ \cite{visser1998,nielsen2005,og2019}:
\begin{align}
ds^2
 = 
\frac{\rho(t, z)}{ c_s(t, z)}
\Big[
& \, - \left(  c_s(t, z)^2- v^{ z}(t, z)^2 \right) dt^2 
 + 2  v^{ z}(t, z) dt d z      \nn \\        
& \, + dx^2 + dy^2 + d z^2
\Big]~.
\label{eq:acmet}
\end{align}
Here, $ c_s(t, z)$ is the speed at which acoustic perturbations 
propagate in the fluid and is defined by $c_s^2=\partial {p}/\partial 
\rho$. Note the relativistic form of the acoustic metric, with speed of 
light c replaced by the sound speed $c_s$, even though the starting fluid 
equations are non-relativistic \cite{visser1998}. We mention here
that our starting fluid equations are for an ideal fluid with zero viscosity. 
Acoustic metrics emerge more naturally in fluids with negligible viscosities
compared to viscous ones, though a derivation of the same for a
relativistic viscous fluid has been given in \cite{bittencourt2018}. Our assumption 
of inviscid nature of the fluid is reasonable in view of relatively small values of the hadronic
phase viscosity \cite{mykhaylova2020,kadam2015}.
Note that this assumption of zero viscosity is unrelated to our earlier 
discussions of ultra relativistic 
collisions where very low values of $\eta/s$ were related to strong quantum 
correlations, justifying quantum nature of the fluid. 


This metric is qualitatively similar to a time dependent, 
spherically symmetric metric written in Painlev\'{e}-Gullstrand coordinates  
\cite{nielsen2005} except that the spatial part in the latter has spherical 
symmetry while ours has axial symmetry about the 
longitudinal (z) axis. At $ z = 0$, 
$ v^z = 0$ and the above metric reduces to that of a flat spacetime. 
An apparent horizon or marginally outer trapped surface forms at a value 
of $ z$ implicitly defined by
\begin{align}
 \lvert v^{ z} (t,  z_H) \rvert =  c_s (t,  z_H)~. 
 \label{eq:achorloc}
\end{align}
The plasma becomes supersonic in the region $ z <  z_H$ 
and any acoustic perturbation from this supersonic region is unable to 
travel upstream against the flow and go \textit{outside} the horizon. 
Surfaces with 
$ z <  z_H$ are said to form (acoustic) outer trapped surfaces and the outermost 
trapped surface is called the marginally outer trapped surface. If we think of a 
spherically symmetric static physical black hole, the marginally outer trapped 
surface coincides with the event horizon and the interior of the event horizon 
forms the trapped region containing outer trapped surfaces. (For a white hole,
we would have similarly had inner trapped surfaces). The acoustic metric 
contains a time and position dependent overall conformal factor, but the 
location of the acoustic horizon as defined here does not depend on it. 

\section{UrQMD analysis for acoustic black hole horizon} \label{sec:urqmd}

\subsection*{Search for local thermal equilibrium} \label{subsec:lte}
UrQMD (Ultra relativistic Quantum Molecular Dynamics) 
is a dynamical transport model  based  on  an  effective  solution  of  the  
relativistic Boltzmann equation used to describe the 
time evolution of a many-body system by using covariant equations of motion. 
The underlying 
degrees of freedom of UrQMD are hadrons (baryons, mesons and their 
antiparticles), their excited states, and resonances. UrQMD also includes  
string  excitation  and  fragmentation (treated according 
to the LUND model \cite{andersson1986,nilsson-almqvist1986,sjostrand1993}),
formation and decay of hadronic resonances as well as rescattering of 
hadrons.   
We use UrQMD-3.3p2 model \cite{bravina1998,bass1998,bleicher1999} to generate 5000 events at 
different times for Au-Au central collisions with laboratory energy of 
$10.7$ GeV (BNL-AGS Experiment)  \cite{barrette1995,barrette1997,barrette1998,barrette2000,liu2000,chung2000,chung2001,pinkenburg2002,pinkenburg2001}.
(We have taken these parameter values for the collisions as sample values 
to demonstrate the possibility of acoustic black hole horizon and 
associated  Hawking radiation in relativistic HIC.)

We start our analysis from time 6 fm (after the collision) by considering 
only nucleons, that is, protons and neutrons. 
In this work we do not include pions because then the resulting fluid would
be better described by relativistic equations (though, as we have mentioned,
still one can argue for the non-relativistic equations to be reasonable
approximation for sub-sonic fluid velocities).  As our formulation is based
on non-relativistic fluid equations, it is much simpler to focus on a nucleon
system which is genuinely a non-relativistic system at the temperatures
considered. (Note that for nucleons, one has to be careful about the initial 
longitudinal velocities, which will not be part of the equilibrium 
distribution. We have taken care of that, as we will see below, by 
fitting Fermi-Dirac distributions at each value of z with
a local fluid velocity which gives completely isotropic velocity 
distribution function.) A hydrodynamic description 
of the system is possible only after local thermodynamic equilibrium (LTE) 
has been attained. To check this, usually one takes a cubic or spherical 
cell around the centre of the system. We begin with a cubic cell of volume 
$5\times5\times5$ fm$^3$ centred around the centre of mass of the system 
located at $z=0$. We follow the procedure of ref. \cite{bravina1998} to check for 
equilibrium in this cell. First we need to verify whether the velocity 
distributions $\frac{dN}{dv^\alpha}$ vs. 
$v^\alpha$ obey  Maxwell Boltzmann (MB) 
velocity distribution given by $\frac{dN}{dv^\alpha} = e^{-m_N \frac{(v^\alpha)^2}{2T}}$ 
and overlap with each other. Here $dN$ is the number of nucleons in the 
velocity bin $dv^\alpha$ with $v^\alpha$ denoting the velocity of individual nucleons 
in $\alpha = x,y,z$ direction, $m_N$ is the mass of a nucleon and $T$ is the 
temperature of the cell. If these distributions overlap with each other, 
that is, if the momenta of the particles are isotropic, local thermal 
equilibrium is possible in this cell. We find that for our choice of cell 
dimension, a time of $6$ fm is too short for achieving local thermal 
equilibrium. In this time, the transverse and longitudinal Maxwell 
Boltzmann velocity distributions do not overlap. In fact,  the longitudinal 
velocity distribution does not even follow a Maxwell Boltzmann distribution 
at $6$ fm time. Full equilibrium can be achieved in such a cell only around 
$13$ fm time.

However, it is not necessary for the cell to be cubic or spherical and we 
find that when we take a rectangular cell with its longitudinal or $z$ sides 
much smaller than the other two, we can get full equilibrium by $6$ fm 
time. This is reasonable as the longitudinal
velocity of particles has extra non-equilibrium contribution from the 
collision geometry, and this component becomes significant away from the
centre of the collision in the longitudinal direction. In particular, when we 
take a cell of  $5\times5\times0.5$ fm$^3$  at $6$ fm time, the Maxwell 
Boltzmann velocity distributions of all velocity components are found to 
overlap \footnote{For example, with a cell of volume $5\times5\times2$ 
fm$^3$, local thermal equilibrium will be achieved at $7$ fm time.}. The 
momentum spectrum of nucleons $\frac{dN}{dp^\alpha}$ vs. $E$, 
with $p^\alpha$ and 
$E$ being the $\alpha^{\text{th}}$ component of three momentum and energy of a 
nucleon respectively, follows a Fermi Dirac distribution, 
$\frac{1}{\exp[(E-\mu_B)/T]+1}$, and by fitting, we obtain the temperature 
of the cell to be $\sim 154$ MeV. Here, $\mu_B$ is the baryonic chemical 
potential. Note that here we are performing low energy collisions and 
taking the cell to be smaller in the $z$ direction. So, baryon density and 
baryon chemical potential are significantly high, especially at 6 fm time. 
This makes quantum statistics significant. In fact, all three overlapping 
velocity distributions show some deviation near the tail from Maxwell 
Boltzmann velocity distribution. For longer times, Boltzmann distribution 
is good enough for fitting \cite{bravina1998}.  

For hydrodynamic description of the system, local thermal equilibrium is 
required not only in the central cell, but in the off-centred cells as well. 
It is not guaranteed that off-centred regions would equilibrate at the same 
time as a central cell if we choose them to be of identical dimensions. By 
allowing the cell shape and size to vary with position, we are able to 
achieve local thermodynamic equilibrium simultaneously in a fairly large part 
of the system, thereby justifying a fluid description of the system.

\subsection*{Locally static acoustic horizon in the plasma} \label{subsec:hor}

\begin{figure}
\centering
\includegraphics[width=0.85\linewidth]{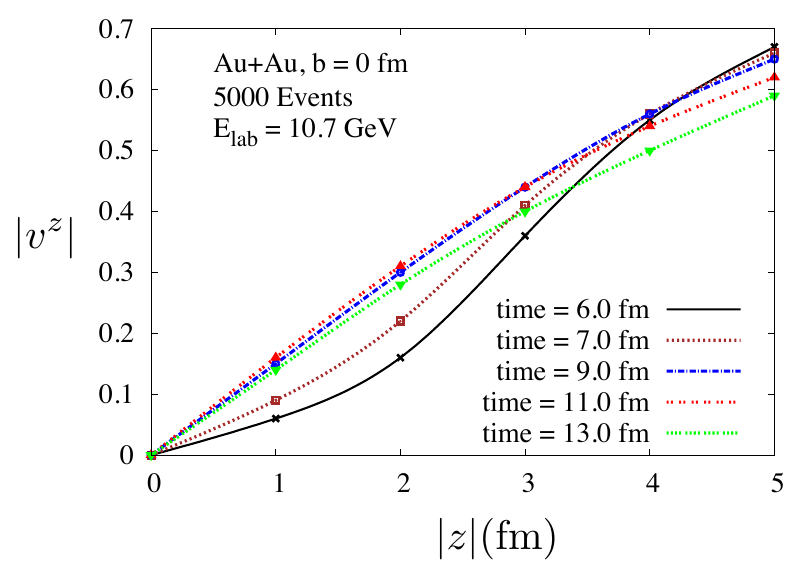}
\caption{\label{fig:vzwithz} Longitudinal velocity of nucleonic fluid 
$\lvert v^z \rvert$ (in units of speed of light, $c$) vs. $\lvert z \rvert$ at different times.  
Plot showing that acoustic horizon 
first moves towards the centre of the plasma, then remains fixed at  
$\lvert z \rvert \approx 2.5$ fm for a time duration of $2$ fm ($9 -11$ fm), 
and then moves outward.}
\end{figure}

In the centre of mass frame, considering both the left and right half spaces, 
the longitudinal Maxwell Boltzmann velocity distribution of the particles, 
$\frac{dN}{dv^z} \, \text{vs.} \, v^z$, in an appropriately chosen cell at 
the origin, is symmetric about $z=0$ while the same in an off-centred cell 
is asymmetric about the value of z corresponding to the centre of that cell. 
If we do a Lorentz transformation in the $z$ direction of longitudinal 
velocities of the particles such that the latter also becomes symmetric, we  
arrive at a frame comoving with the fluid at that specific value of $z$. 
This gives us the velocity of the fluid at that 
particular $z$, as measured in the centre of mass frame. 
We have performed calculation of plasma velocity 
for the region where $z \geq 0$.
Due to symmetry of collision about $z=0$, similar behaviour should
 hold for $z \leq 0$ region also.
In this way, we can 
find the velocity of the fluid at different points in the region 
$0 < \lvert z \rvert < 5$ fm at different times. The result is plotted in 
figure \ref{fig:vzwithz} and shows the following features:
\begin{enumerate}
	\item for $0<\lvert z \rvert<4$ fm, the general behaviour is that at a 
	fixed $z$, $\lvert v^z \rvert$ initially increases with time and then starts 
	decreasing. The time at which this transition occurs depends on $z$, 
	happening earlier at higher values of $\lvert z\rvert$;
	\item for $4 \text{fm} <\lvert z\rvert <5 \text{fm}$, $\lvert v^z 
	\rvert$ is monotonically decreasing.
\end{enumerate}
This behaviour cannot be explained if we na\"ively assume Bjorken expansion 
\cite{bjorken1982} for the plasma where $v^z \sim z/t$ \cite{bjorken1982}.
However, as we discussed above, for low energy collision with large nuclear 
stopping, Bjorken picture is not appropriate \cite{bjorken1982}. In our 
simulations, 
the centre of mass energy of the collision is low. This 
results in a narrow rapidity distribution which in turn gives rise to a 
strong pressure gradient along the $z$ direction in the plasma. This 
pressure gradient accelerates the plasma in the region $0<\lvert z 
\rvert<4$ fm but dies off near $\lvert z\rvert = 4$ fm and beyond that, 
the longitudinal velocity at any fixed $\lvert z\rvert$ decreases with 
time owing to expansion. 

In ref. \cite{bravina1998}, the authors determine the equation of state for a 
hadronic system produced by the same collision occurring at the same energy 
as considered here, in the time interval $10$ fm$-18$ fm to be $p \simeq 
0.12 \epsilon$ in the central cell. Note, that the equation of state does 
not vary during this interval. Thus, the hadronic system has a constant 
speed of sound given by $c_s = 0.35$ (in units of $c$) in this period. If we assume that the 
same equation of state remains valid in the preceding time interval of 
$6 \ \text{fm} - 11 \ \text{fm}$ and is applicable to off-centred cells too, 
then an acoustic horizon is formed in the plasma at the position where
\begin{align}
\lvert v^z (t, z_H)\rvert = 0.35.
\end{align}
Figure \ref{fig:vzwithz} reveals an interesting behaviour of this acoustic 
horizon - in the beginning, from time $0 \ \text{fm} - 9 \ \text{fm}$, it moves 
towards lower values of $\lvert z \rvert$, then remains static at  $\lvert 
z \rvert \simeq 2.5$ fm for a duration of about $2$ fm and again starts moving 
after $t \approx 11$ fm, but now towards higher values of $\lvert z \rvert$. 
Even if the speed of sound in the plasma has some other constant value, we 
would get a similar behaviour of the horizon as long as $0<c_s<0.58$ (beyond
this value, there is no time period over which one gets static horizon,
in figure \ref{fig:vzwithz}). Though, for low energy collision, it is improbable that $c_s$ 
would reach  $0.58$. Of 
course, the associated time intervals in such situations would be different. 
However, if the speed of sound varied temporally and spatially, it would not 
be possible to predict the evolution of the horizon without explicitly 
knowing $c_s(t,z)$. On the other hand, if we had considered high energy 
collisions, like those occurring at RHIC and LHC, a strong pressure gradient 
would occur only for values of $z$ corresponding to the decaying part of 
the rapidity distribution. In such situations, the speed of sound in the 
plasma would also increase to about $0.58$. So, a similar evolution of the 
acoustic horizon could be expected.

\section{Surface gravity and Hawking temperature} \label{sec:sgrhr}

In figure \ref{fig:vzwithz}, the curves corresponding to $t=9$ fm and 
$t=11$ fm almost overlap upto $\lvert z \rvert \approx 4$ fm. For $c_s \approx 0.35$, 
the acoustic horizon lies close to $\lvert z \rvert \approx 2.5$ fm 
during this time interval. Remember that we have decided to look at the 
analogue spacetime emerging only in the plasma occupying the left half space 
(see figure \ref{fig:vzwithz}). Regions of the acoustic spacetime with 
$ z < - 2.5$ fm lie inside the acoustic black hole. Hawking radiation 
occurs from the part of spacetime lying just outside the horizon, but to 
experimentally identify signatures of this radiation, it may be necessary 
for us to probe the region lying inside the horizon too. Otherwise, from 
the point of view of a gravitational analogue, it is only the spacetime 
outside the horizon that is of concern to us. Now, an almost time independent 
profile of $v^z$ in the interval $9 \ \text{fm} < t < 11 \ \text{fm}$ for 
$-4 \ \text{fm} <  z < 0$ offers us the great 
advantage of approximating the general dynamical metric of eq. \eqref{eq:acmet} 
as a conformally static one given by:
\begin{alignat}{2}
 g
& \, = 
\frac{\rho(t, z)}{ c_s}
\Big[
&& \, - \left(  c_s^2 -  v^{ z}( z)^2 \right) dt \otimes dt
+ 2  v^{ z}( z) ~dt \otimes d z   \nn \\      
& \, && \,  + 
dx\otimes dx + dy\otimes dy + d z \otimes d z
\Big]~
\label{eq:g}	 \\
& \, = \Omega(t, z) ~\tl g
\label{eq:gtlg}
\end{alignat}
where $\Omega(t, z) = \frac{\rho(t, z)}{ c_s}$ is a conformal 
factor depending on position in the acoustic spacetime. At first, we ignore 
the conformal factor. This brings us to a static acoustic metric 
$\tl g$ that is 
qualitatively similar to a static Schwarzschild metric written in 
Painlev\'{e}-Gullstrand coordinates with axial
symmetry instead of spherical symmetry. It has a timelike Killing 
vector field $\tl k^a = (1,0,0,0)$ 
\footnote{All quantities defined in the spacetime with metric $\tl g$ 
will be distinguished by a tilde on top.}. 
This normalisation naturally gives 
\begin{align*}
\tl g(\tl k,\tl k)= - ( c_s^2 -  v^{ z}( z)^2)
\end{align*} 
which gives $\tl g(\tl k,\tl k) = - c_s^2$ at $ z = 0$ ensuring that $\tl k$ matches 
with the four velocity of an observer in the asymptotically flat region of the acoustic 
spacetime 
(note the difference with general relativity where the speed of light is set to unity). 
The apparent horizon 
defined by $ c_s( z_H) =  \lvert v^{ z}( z_H) \rvert$ is also a Killing 
horizon for $\tl k$ since $\tl g(\tl k,\tl k)=0$ on this surface. The surface gravity 
$\tl \kappa$ of a Killing horizon is defined by the relation
\begin{align}
\nabla_a \left(\tl g_{bc} \tl k^b \tl k^c \right)
= - 2 \tl\kappa \tl g_{ab} \tl k^b .
\label{eq:sgr}
\end{align}
For the metric $\tl g$, a simple calculation gives 
%
\begin{align*}
\tl \kappa = \left( c_s' - ( v^{ z})^\prime \right) \big\vert_H
\label{eq:tlkappag}
\end{align*} 
and as $c_s$ is constant  in our model, we simply have 
\begin{align}
\tl\kappa = - ( v^{ z})^\prime \vert_H.
\end{align} 
Since, $( v^{ z})^\prime \vert_H$ 
is negative, $\tl\kappa$ remains a positive quantity. Now, under the conformal 
transformation 
\begin{align}
\tl g (t, z) \to  g (t, z) = \Omega (t, z)~\tl g(t, z),
\end{align}
$\tl k$ will remain a Killing vector field if  
\begin{align}
\tl k(\Omega) = \tl k^0 \partial_0 \Omega = 0.
\end{align}
Otherwise, $\tl k$ becomes a conformal Killing vector field and the Killing 
horizon a conformal Killing horizon:
$g (\tl k,\tl k) \vert_H = \Omega\vert_H \tl g(\tl k,\tl k)\vert_H = 0$. 
The definition of surface gravity by eq. \eqref{eq:sgr} is conformally invariant 
even if $\tl k$ doesn't remain a true Killing vector field because, on the 
(conformal) Killing horizon,
\begin{align}
\nabla_a (g_{bc} \tl k^b \tl k^c)
& \, =
\nabla_a (\Omega \tl g_{bc} \tl k^b \tl k^c) \nn \\
& \, = \tl g_{bc} \tl k^b \tl k^c \nabla_a \Omega + \Omega \nabla_a(\tl g_{bc} \tl k^b \tl k^c) \nn \\
& \, =
- 2 \Omega\tl\kappa \tl g_{ab} \tl k^b \nn \\
& \, = - 2 \tl\kappa g_{ab} \tl k^b.
\end{align}
The first term in the second equality vanishes because on the Killing horizon 
$\tl g(\tl k,\tl k)=0$. However, for $\tl\kappa$ to be the surface gravity measured by an 
asymptotic observer in a spacetime with metric $g$, the normalisation 
of $\tl k$ has to be appropriately changed: $\tl k^a \to k^a = 
\frac{1}{\sqrt{\Omega(t,0)}} \tl k^a$ such that 
\begin{align}
g(k, k) \vert_{z=0} 
= \Omega (t,0) \ \tl g(k,k)\vert_{z=0} 
= \tl g(\tl k,\tl k)\vert_{z=0} 
= - c_s^2.
\label{eq:knorm}
\end{align} 
The new surface gravity $\kappa$ is obtained using the defining relation
\begin{align}
\nabla_a(g_{bc} k^b k^c)
= - 2 \kappa g_{ab} k^b.
\end{align}
This makes 
\begin{align}
\kappa = \frac{\tl\kappa}{\sqrt{\Omega(t,0)}}
\end{align} 
and it is the surface gravity measured by an observer in the 
conformally flat asymptotic region of the acoustic spacetime with metric $g$. 
Such an observer sees a thermal spectrum of spontaneous radiation, the 
Hawking radiation, emanating from the (conformal) Killing horizon and 
when the corresponding (conformal) Killing vector field is normalised such 
that it matches with her four velocity in the asymptotic region (as done 
in eq. \eqref{eq:knorm}), she finds that this radiation is at a temperature \cite{nielsen2012}
\begin{align}
T
= \frac{\kappa}{2\pi}
= \frac{\tl \kappa}{2\pi \sqrt{\Omega (t,0)}}
= \frac{\tl T}{\sqrt{\Omega (t,0)}}.
\label{eq:ht0}
\end{align}
Here, $\tl T$ would have been the temperature measured by her if she 
were in the spacetime with metric $\tl g$. 

We now note that the acoustic metric in eq. \eqref{eq:acmet} is derived 
starting from (non-relativistic) fluid equations that allow the freedom to 
multiply the metric by an overall constant. We utilize this to replace the 
conformal factor $\Omega(t,z)$ in eq. \eqref{eq:g} by 
$\Omega(t,z)/\Omega (t_0,0)$.
We take $t_0$ to be some value in the time interval $(9-11)$ fm relevant
for our discussion in this section. 
Thus, in the asymptotically flat region, at $z=0$, the new normalised 
conformal factor becomes unity at $t=t_0$ 
and the temperature, as measured by the asymptotic observer, becomes
\begin{align}
k_B T = - \frac{\hbar}{2\pi} (v^z)^\prime \big\vert_{H}    .
\label{eq:htsgr}
\end{align}
Here, we have restored the fundamental constants $\hbar,k_B$ in the 
expression for Hawking temperature. Over the time period 
$9 \ \text{fm} < t <11 \ \text{fm}$, the central density
actually decreases. Asymptotic value of the conformal factor then does not
remain $1$, changing to ${\Omega(t,0) \over \Omega(t_0,0)} < 1$ for $t > t_0$.
For 1-d expansion, central density decreases linearly, so for $t_0 \sim
10$ fm, the change in density is about 20 \% in a time duration
of 2 fm ($t \sim (9-11)$ fm). The proper time for the asymptotic observer
is then changed by the factor $\sqrt{\Omega(t,0)/\Omega(t_0,0)}$, which is
about 10\% (for density change of order 20\%). The temperatures measured
by the asymptotic observer are then blue shifted by this amount.
Thus, we conclude that the temperature observed by the asymptotic
observer (at $z = 0$) is given by eq. \eqref{eq:htsgr}, with the value possibly 
increasing by about 10\% due to decreasing central density.
(Even though we have avoided consideration of dynamical horizon here, it is
tempting to point out that for earlier time period ($t < 9$ fm), the sonic
horizon actually moves towards $z = 0$. One should then expect blue shifted
Hawking radiation as observed by the asymptotic observer at $z = 0$.)

Before we discuss estimate of this temperature, it is instructive to
derive this temperature using imaginary time formalism.
For this we start with the following expression for the acoustic metric in $(\tau,x,y,z)$
coordinates:
\begin{align}
ds^2
= 
\Omega(\tau,z)
\Bigg[
& \, - \left(  c_s^2- v^{ z}(\tau, z)^2 \right) d\tau^2  \nn \\
& \, + \left( \delta_{\alpha\beta} + \frac{v_\alpha v_\beta}{c_s^2 - v^2} \right) dx^\alpha dx^\beta
\Bigg]~.
\end{align}
 Here, $v_\alpha = \delta_{\alpha\beta} v^\beta$. This metric is related  to the one in eq. \eqref{eq:g} by a coordinate 
 transformation from $t \rightarrow \tau$ \cite{visser1998}:
\begin{equation}
	d\tau = dt + {{\vec v}.{\vec dx} \over c_s^2 - v^2}
\end{equation}
with velocity along z-axis.  (Note that in the conformal factor, the
density now depends on the new variable $\tau$.) We  get,
\begin{align}
ds^2
 = 
\Omega(\tau,z)
\Big[
	- c_s^2 d\tau^2 f(z) 
	 + {dz^2 \over f(z)} + dx^2 + dy^2 
\Big]
\end{align}
 where
\begin{equation}
f(z) = 1- \frac{v^z(z)^2}{c_s^2}. 
\end{equation}
For the near horizon geometry, we expand $f(z)$ about $z = z_H$
for $z > z_H$,
\begin{equation} 
f(z) = \frac{\partial f(z)}{\partial z}\bigg\vert_{z_H} (z-z_H)
\end{equation}
using $f(z_H) = 0$ as $v^z (z_H) = c_s$. We now define a coordinate $\rho_{sp}$
which is the proper distance from the horizon (subscript {\it `sp'} is used
to distinguish from the symbol $\rho$ used for density):
\begin{equation}
d\rho_{sp} = \frac{dz}{\sqrt{f(z)}} 
= \frac{dz}{\sqrt{f^\prime(z_H)} \sqrt{z-z_H}}.
\end{equation}
Integration gives
\begin{equation}
	\rho_{sp} = 2 \frac{\sqrt{z-z_H}}{\sqrt{f^\prime(z_H)}}.
\end{equation}
With this we can express $f(z)$ as a function of $\rho_{sp}$
\begin{equation}
f(z) = \frac{\rho_{sp}^2}{4} f^\prime(z_H)^2  \equiv f(\rho_{sp}).
\end{equation}
The near horizon metric then becomes
\begin{equation}
	ds^2 = \Omega(\tau,z) \left[-K^2 \rho_{sp}^2 d\tau^2 
	+ d\rho_{sp}^2 + dx^2 + dy^2 \right], 
\end{equation}
where 
\begin{equation}
K = {f^\prime(z_H) c_s \over 2} ~ = - (v^z)^\prime(z_H),
\end{equation}
as $v^z(z_H) = c_s$.  Rest of the procedure is 
standard. We go to the Euclidean space with $\tau \rightarrow -i\tau_E$:
\begin{align}
	ds^2 
	& \, = \Omega(\tau,z) \Big[K^2 \rho_{sp}^2 d\tau_E^2 
	+ d\rho_{sp}^2 + dx^2 + dy^2 \Big]  \nn \\
	& \, = \Omega(\tau,z) \Big[\rho_{sp}^2 d\theta^2 
	+ d\rho_{sp}^2 + dx^2 + dy^2 \Big],
\end{align}
where $\theta = K\tau_E$. The two dimensional space spanned by
$(\rho_{sp},\theta)$ has a
conical singularity at $\rho_{sp}=0$ i.e. the location of the horizon,
unless $\theta$ is periodic with a period of $2\pi$. Since, there is no 
physical singularity at the horizon, we must have $\theta \sim \theta + 2\pi$.
This implies that $\tau_E$ is also periodic:
\begin{equation}
\tau_E \sim \tau_E + {2\pi \over K}.
\end{equation}
We thus conclude that the quantum fields in this space-time (which
are fluctuations in the velocity potential) will be in thermal equilibrium
with a temperature $T$,
\begin{equation}
T = \frac{K}{2\pi} = - \frac{(v^z)^\prime(z_H)}{2\pi}
\end{equation}
which is the same as obtained earlier in eq. \eqref{eq:htsgr}.
The remaining argument is same as that following eq. \eqref{eq:htsgr}. Suitably 
normalising by the factor $1/\Omega(\tau_0,0)$, we conclude that the temperature
observed by the asymptotic observer at ($z = 0$) is given by the above
equation.  (It is important to note that the conformal factor now depends on
$\tau$, so the value $\tau_0$ will correspond to
$t = t_0$ at $z = 0$.)
Further, following the same argument about the effect of decrease in the 
central density on the conformal factor (as after eq. \eqref{eq:htsgr}), we
conclude that the observed temperature may be larger by about 10\%
(when density decreases by about 20\% for the relevant duration of time).

 We now estimate the value of temperature. The relevant parts of the plots
 in figure \ref{fig:vzwithz} lie between $\lvert z \rvert = 2-3$ fm. 
 Recall that we had chosen to confine our attention to the region
  $z < 0$. The horizontal axis then should be
 read as negative values of $z$. To calculate $(v^z)^\prime(z_H)$ we use
 plots for $t = 9$ and $t=11$ fm between $z = -3$ and $z= - 2$ fm, as this
 encloses the value $v^z(z_H) = c_s = 0.35$. We get $(v^z)^\prime(z_H) 
 \simeq 0.12$ fm$^{-1}$. The value of temperature is then
\begin{equation}
k_B T = \hbar \frac{(v^z)^\prime(z_H)}{2\pi} \simeq \frac{0.12}{2\pi} ~\text{fm}^{-1}
\simeq 4  ~\text{MeV} .
\end{equation}
The temperature can increase by about 10 \%  due to decrease
in central density by about 20\% for the relevant time period.
(We again mention that for $t < 9$ fm, the sonic horizon actually moves
towards the asymptotic observer at $z = 0$. From general physical
consideration, one will expect higher value of the Hawking temperature
due to the blue shift of the radiation.)

\section{Observational prospects and conclusion} \label{sec:obs}

 As the temperature of the plasma, even at low collision
 energies considered here,  is about 135 MeV, it may be difficult to 
 observe the signature of Hawking temperature of about 4-5 MeV.
 However,  note that here the Hawking radiation will be 
 from the region of acoustic horizon which will be away
 from the central region (where the plasma temperature is maximum). The
 expected temperature for large rapidity regions is smaller,
 which may help in identifying the signal here. For observations, we recall
that the temperature of the plasma at the freezeout surface 
depends on the longitudinal coordinate z, primarily due to z dependence of
the chemical potential and also due to departure from strict Bjorken
longitudinal boost scaling. Further, the longitudinal flow velocity of
the plasma depends on z coordinate. Thus, hadrons coming out from a
specific z value have specific rapidity contribution from this flow
(called as frame rapidity) and also correspond to local z dependent
temperature. This leads to well defined rapidity dependent particle
distributions.

In the presence of Hawking radiation, this rapidity dependence of particle
distributions will be changed. This is because the relevant scalar field here
is the scalar perturbation over velocity potential, $\psi$,  and 
hence, the Hawking radiation
appears in the form of a thermal spectrum of excitations of $\psi$. Thus, at
any value of z (between the collision centre and the acoustic black hole
horizon) there will be a thermal distribution of 
perturbations in the velocity potential giving rise to
 a fluctuating component in flow velocity at that specific z value.
Thus, the earlier z dependent flow velocity (in the absence of Hawking
radiation) gets a fluctuating component. Also, the z dependence of temperature
gets affected due to fluctuations in the local velocity. This directly
affects the rapidity dependence of particle distributions. 
Specifically, say, rapidity dependence of transverse momentum $p_T$ 
distributions of particles should have a thermal component. Thus, while 
fitting the observed rapidity dependence for these distributions, one should
allow a thermal part in the rapidity distribution. We have not worked
out the specific dependence yet coming from Hawking radiation, it requires
detailed hydrodynamic simulations. We hope to present it in a future work.
Our purpose in the present work has been to illustrate the 
basic physics of the 
phenomenon, and demonstrate its existence for our system under
 specific conditions.

  For the present work, we have avoided the issue of dynamical horizon
  so that conceptual issues do not overshadow the main physics we want to
  illustrate, the possibility of observing Hawking radiation from acoustic
  black holes in heavy-ion collisions. Due to this limited focus, we have
  found a narrow window of time $t$ of about $9 -11$ 
  fm during which static horizon
  could be achieved. If freezeout happens much later, then this thermal
  component may get lost in subsequent thermalisation. In such a situation,
  one may be able to observe this signal in terms of thermal photons and
  di-leptons which come out of the QGP region without scattering. Rapidity
  dependence of distribution of these particles should then contain
  a hidden signal of a thermal stage for the rapidity variable, even for a 
  short period of time.

 As we discussed above, we restricted to low energy collisions because
 here static horizon could be achieved rather easily, even without 
 considering possible variations in the sound velocity. As the Hawking
 temperature is related to the velocity gradient, for ultra-relativistic
 collisions, where thermal equilibrium is reached very quickly (in times
 much less than 1 fm for LHC energies), the velocity gradient will be
 far larger, leading to large Hawking temperature. It is very unlikely
 to get static acoustic horizon in those cases, however, due to non-trivial
 energy density/pressure gradients, it may be possible to make the velocity 
 of acoustic horizon much smaller than the sound velocity by suitable 
 choices of nuclei and collision energies. 
 With proper consideration of dynamical
 horizons then, a large Hawking temperature may be possible in those cases.
 Relevant fluid equations will then be relativistic hydrodynamics equations.
 Acoustic metric for relativistic hydrodynamics has been discussed in the
 literature \cite{visser2010,bilic1999,ge2010}, and we plan to discuss this very exciting possibility in future.

\paragraph{Acknowledgment}
We are  grateful to Amitabh Virmani, Sanatan Digal, Ananta P. Mishra, 
Ranjita K. Mohapatra, Yogesh Srivastava, Raghu Rangarajan, Nirupam Dutta 
and Rajiv V. Gavai for useful comments and discussions. We would like to 
specially thank Somnath De for introducing UrQMD to us. A part of the work done 
by AD is supported by the Polish National Science Center Grant 
No. 2018/30/E/ST2/00432. OG sincerely thanks 
Institute of Physics, Bhubaneswar for financial support and hospitality as 
this work was conceived and semi-completed during an extended visit there. 
We also thank an anonymous referee for a very careful checking of the paper
and for making very important suggestions and corrections.

\section*{Bibliography}

\bibliographystyle{elsarticle-num}
\bibliography{gravity,an_gr,an_bec,qgp,dynhor,opub}

\end{document}